\documentclass{article}

\usepackage{microtype}
\usepackage{graphicx}
\usepackage{subfigure}
\usepackage{booktabs}
\usepackage{hyperref}

\usepackage{mlsys2025}
\usepackage{amsmath,amsfonts,amssymb,xspace,xcolor,enumitem}
\usepackage{tabularx,array,multirow,makecell}
\usepackage{siunitx,float,placeins}

\graphicspath{{figures/}}
\newcommand{\sys}{\textsc{MARS}\xspace}

\makeatletter
\gdef\isaccepted{1}
\renewcommand{\Notice@String}{Preprint. Under review.}
\makeatother

\hypersetup{
  pdfauthor={Yifei Wang, Tsun-yu Yang, Hancheng Ye, Yechen Xu, Cong Guo, Chiyue Wei, Qinsi Wang, Dongting Li, Tingjun Chen, Hai Helen Li, Danyang Zhuo, Yiran Chen},
  pdfsubject={Public research preprint},
  pdfkeywords={LLM agent serving, KV cache thrashing, workflow goodput, GPU memory management},
}

\mlsystitlerunning{MARS: Towards High Goodput and Low KV Thrashing in Agent Serving under GPU Memory Constraints}

\begin{document}
\twocolumn[
\mlsystitle{MARS: Towards High Goodput and Low KV Thrashing in \\ Agent Serving under GPU Memory Constraints}

\begin{mlsysauthorlist}
  \mlsysauthor{Yifei Wang}{duke}\hspace{0.35em}
  \mlsysauthor{Tsun-yu Yang}{duke}\hspace{0.35em}
  \mlsysauthor{Hancheng Ye}{duke}\hspace{0.35em}
  \mlsysauthor{Yechen Xu}{duke}\hspace{0.35em}
  \mlsysauthor{Cong Guo}{duke}\hspace{0.35em}
  \mlsysauthor{Chiyue Wei}{duke}\\
  \mlsysauthor{Qinsi Wang}{duke}\hspace{0.35em}
  \mlsysauthor{Dongting Li}{duke}\hspace{0.35em}
  \mlsysauthor{Tingjun Chen}{duke}\hspace{0.35em}
  \mlsysauthor{Hai ``Helen'' Li}{duke}\hspace{0.35em}
  \mlsysauthor{Danyang Zhuo}{duke}\hspace{0.35em}
  \mlsysauthor{Yiran Chen}{duke}
\end{mlsysauthorlist}
\mlsysaffiliation{duke}{Duke University, Durham, North Carolina, USA}
\vskip 0.18in
\begin{abstract}

Self-hosted LLM agents offer greater control over sensitive data and model customization while avoiding per-request API fees. However, serving concurrent, multi-turn workflows within limited GPU memory makes it difficult to achieve high workflow-level goodput. In particular, as workflows accumulate context, their reusable key-value (KV) caches can easily outgrow GPU capacity, triggering repeated eviction and recovery, known as KV cache thrashing. This process consumes resources and delays workflow completion. Although existing approaches propose various methods, our motivation study shows that the evaluated baselines still incur substantial thrashing overhead or fail to address the workflow-level goodput issue.

We present \sys, an agent serving system that coordinates workflow admission and in-engine scheduling to achieve high workflow-level goodput on memory-constrained devices while avoiding KV cache thrashing.
Specifically, outside the engine, an admission controller selects root workflows using initial prefix footprint estimates and live KV availability, adapting admissions to memory pressure. Inside, an agent-centric scheduler prioritizes turn completion and coordinates prefill allocation with KV reclamation to reduce disruptive evictions. Across two models and three GPU configurations, \sys achieves $1.59\times$--$7.56\times$ mean end-to-end latency speedups over the fastest baseline. Integrated as the serving backend for OpenCode, Codex, and OpenHands, \sys further improves mean workflow completion time by $1.17\times$--$1.87\times$.

Our source code is publicly available at \url{https://github.com/Afterglow231/MARS_preview}.

\end{abstract}

\vskip 0.12in
]

% First-page affiliation footnote, matching the Tideway preprint layout.
% Do not invent corresponding author email addresses.
\footnotetext[1]{Duke University, Durham, North Carolina, USA.\par\smallskip Preprint. Under review.}
\hypersetup{pdfauthor={Yifei Wang, Tsun-yu Yang, Hancheng Ye, Yechen Xu, Cong Guo, Chiyue Wei, Qinsi Wang, Dongting Li, Tingjun Chen, Hai Helen Li, Danyang Zhuo, Yiran Chen}}

\section{Introduction}

Autonomous agents are shifting large language model (LLM) serving from single-turn queries to multi-turn workflows. Across orchestration frameworks~\cite{hong2024metagpt,wu2024autogen,chase2022langchain} and applications in software engineering~\cite{yang2024sweagent,wang2025openhands}, deep research~\cite{openaiDeepResearch2025}, and computer use~\cite{xie2024osworld}, agents iteratively plan and invoke tools until a task completes. While many agent applications rely on cloud-hosted LLM services, self-hosted deployment is an attractive alternative for many companies and organizations who want to preserve data privacy, avoid API fees, and customize models for specialized domains~\cite{song2024powerinfer,erdogan2024tinyagent}.

For local deployment, serving concurrent agent workflows effectively requires both a high completion rate (known as goodput) and acceptable end-to-end latency.

However, in memory-constrained deployments, concurrent agent workflows make this goal challenging.
Specifically, each agent workflow accumulates model outputs and tool results across turns, allowing subsequent LLM requests to reuse the KV cache of previously processed tokens. As contexts grow and child sessions appear, their combined KV demand easily exceeds GPU capacity.
To accommodate other requests, the engine may evict an ongoing workflow’s reusable state, which must be recomputed or reloaded when that workflow resumes. This repeated eviction and recovery, known as \textit{KV cache thrashing}, consumes resources and delays successive inference turns, making workflows more likely to miss their end-to-end latency budgets. Yet, avoiding thrashing alone does not ensure high goodput. Even when reusable state remains available, scheduling policies can still slow individual turns and delay workflow completion despite high aggregate throughput. Therefore, how to avoid KV thrashing while achieving high goodput becomes the key challenge in memory-constrained agent serving.

Existing scheduling and caching approaches fall short of achieving both objectives under limited memory.
For example, Autellix~\cite{autellix} approximates shortest-job-first (SJF) via program-level attained service but leaves KV demand unbounded, triggering KV thrashing frequently.
Other approaches focus on reducing thrashing through local KV cache heuristics.
InferCept~\cite{infercept} uses a cost function to trade off recomputation against swapping, while LightLLM~\cite{gong2025past} gates requests using predicted peak token lengths. But both of them struggle when multi-turn contexts grow unpredictably.
Continuum~\cite{continuum} uses time-to-live (TTL) policies to further improve KV retention across tool intervals, but still incurs non-negligible recovery overhead once active KV demand exceeds GPU capacity. Moreover, these cache-centric methods overlook workflow-level prioritization, thus frequently missing latency budgets and yielding low goodput.

Our \emph{key observation} is that existing systems fail to reconcile both objectives as they build on throughput-oriented engines (e.g., vLLM~\cite{kwon2023vllm}) that favor \emph{large batches to maximize token throughput}.
However, under agent workloads, large-batch execution can reduce goodput for two reasons. First, large batches increase concurrent KV memory demand, leaving less GPU memory to preserve reusable prefixes across turns. Under memory pressure, these prefixes may be evicted and later recomputed or reloaded when workflows resume, triggering KV thrashing and delaying workflow progress. Second, sharing GPU resources across a large batch can slow individual decoding turns, even when aggregate token throughput remains high. These delays accumulate across successive turns, causing workflows to miss their end-to-end SLOs and reducing goodput.

Our \emph{key insight} is that batch size should be appropriately chosen to improve workflow-level goodput under limited GPU memory. This requires sufficient parallelism for efficient GPU execution while limiting KV contention and allowing prioritized calls to complete promptly. However, \textit{the appropriate batch size alone is insufficient}, as workflows that initially fit in GPU memory may outgrow its capacity as their contexts grow across turns. Therefore, the system must coordinate which workflows enter the engine with how their calls execute and consume KV memory over time. This motivates a unified serving design that combines memory-aware workflow admission with in-engine scheduling and KV cache management.

To this end, we present MARS, an agent serving system that coordinates workflow admission and in-engine scheduling to achieve high workflow-level goodput on memory-constrained devices while avoiding KV cache thrashing.
\sys shapes the engine-visible working set and coordinates two components across the engine boundary. Outside the engine, an \emph{Admission Controller} regulates workflow admission using initial prefix footprint estimates and live KV availability. It favors workflows with smaller initial footprints under contention and adapts the admission window to context growth and memory pressure.

Inside the engine, an \emph{Agent-Centric Scheduler} prioritizes turn completion and coordinates prefill allocation with KV reclamation to reduce disruptive evictions. It adjusts per-call priority using prefill demand and runtime feedback, resizes per-iteration prefill chunks to fit available KV blocks, and applies priority-aware preemption when memory is exhausted. In effect, \sys shifts serving from \emph{evict-and-recover} to \emph{admit-and-finish}, sustaining timely workflow completion while maintaining comparable token throughput under memory constraints.

We implement \sys on top of vLLM~\cite{kwon2023vllm} and integrate it with OpenHands~\cite{wang2025openhands}, OpenCode~\cite{opencode2025} and Codex CLI~\cite{openaiCodexPrompting2026}. Across H100, H200, and A5000 configurations, Qwen3-Coder-30B~\cite{qwen3} and GPT-OSS-120B~\cite{openai_gpt_oss_2025}, and workloads from SWE-bench~\cite{jimenez2024swebench}, GitTaskBench~\cite{ni2025gittaskbench}, Terminal-Bench~\cite{merrill2026terminalbench}, RepoBench~\cite{liu2024repobench}, and $\infty$Bench~\cite{zhang2024inftybench}, \sys achieves $1.59\times$--$7.56\times$ speedups in mean latency over the fastest baseline. In full-stack deployments with agent frontends, \sys further improves mean completion time by 1.17$\times$--1.87$\times$.

\section{Motivation}
\label{sec:motivation}

\begin{figure}[t]
    \centering
    \includegraphics[width=\linewidth]{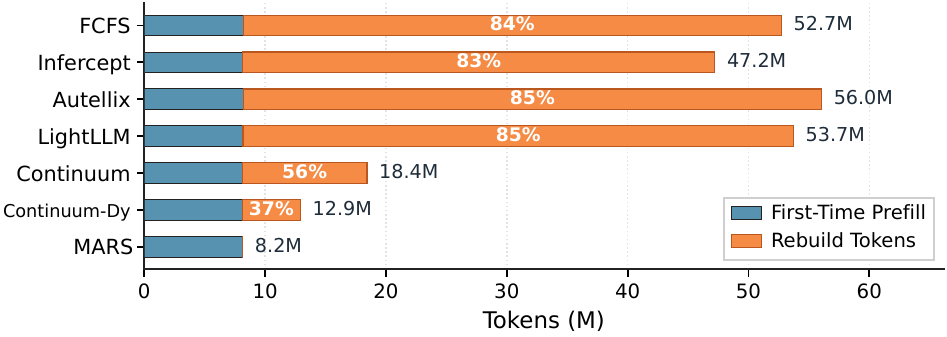}
    \vspace{-8mm}
    \caption{Rebuild work across baselines. Percentages denote the fraction of prefix tokens caused
    by rebuilds.}
    \label{fig:rebuild}
    \vspace{-3mm}
\end{figure}

When agent workflows share a memory-constrained server, their accumulating context prefixes quickly push the active KV working set beyond GPU memory capacity. In this section, we first quantify the prefix rebuild overhead across existing systems (\S\ref{sec:rebuild_overhead}). We then introduce a workflow-level metric, \emph{Per-Task SLO-Aware Goodput}, to reflect end-to-end SLO attainment (\S\ref{sec:goodput}).
Finally, we give a breakdown of execution time per token across baselines (\S\ref{sec:break_down}).

\subsection{Overhead from Prefix Rebuilds}
\label{sec:rebuild_overhead}

Existing serving systems remain vulnerable to engine-level thrashing as they ignore incoming requests' KV demand and live hardware memory pressure. When admission and execution expose more KV demand than the GPU can sustain, new allocations evict or offload reusable prefixes from ongoing workflows (e.g., those waiting for tools) to free space.

When those workflows return for subsequent turns, the engine is forced to reconstruct their discarded KV states before decoding can resume, which we call \textit{prefix rebuilds}.

To quantify this overhead, we decompose prefix processing into \emph{first-time prefill}, which computes KV state for newly introduced input tokens and tool outputs, and \emph{prefix rebuilds}, which recompute previously processed prefixes after their KV state is evicted.
Figure~\ref{fig:rebuild} shows this breakdown for the ILR-1 workload (\S~\ref{sec:workload-analysis}). All systems process the same 8.2M first-time prefill tokens. The baselines, however, process 12.9M--56.0M prefix tokens in total, with rebuilds accounting for 37\%--85\% of this work. By bounding the admitted KV working set within GPU memory, \sys eliminates nearly all prefix rebuilds.

\begin{figure}[t]
    \centering
    \includegraphics[width=0.92\linewidth]{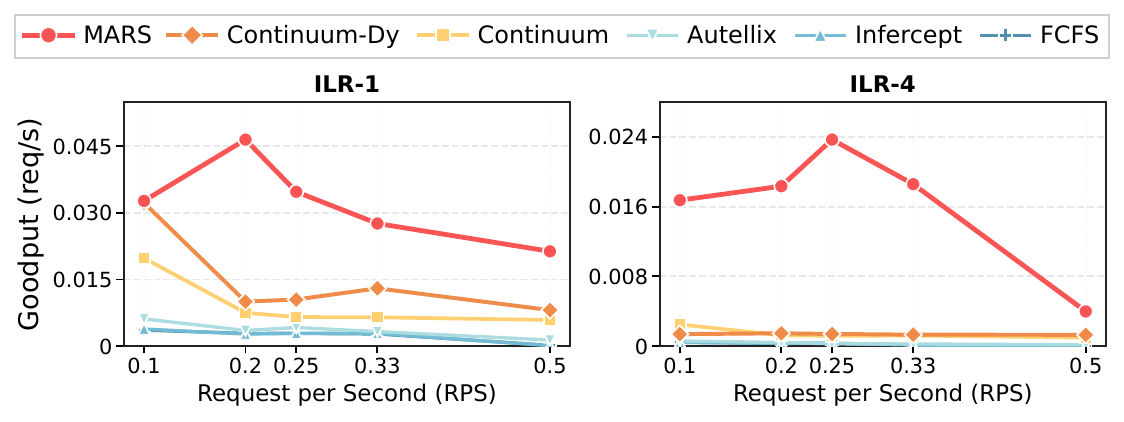}
    \vspace{-3mm}

    \caption{Workflow-level SLO-aware goodput on ILR-1 and ILR-4 at $\alpha=3$. \sys sustains higher goodput than all baselines. }
    \label{fig:goodput_collapse}
    \vspace{-3mm}
\end{figure}

\subsection{Workflow-Level SLO-Aware Goodput}
\label{sec:goodput}

This section examines whether existing systems can complete concurrent workflows within their latency budgets.

To this end, we introduce \textit{Per-Task SLO-Aware Goodput} to measure the workflows' completion rate within an $\alpha$-fold slowdown of their isolated execution time.
Formally, let $\mathcal{W}_{\Delta t}$ denote the set of workflows completed within a time window $\Delta t$. Each workflow $i$ has an end-to-end latency $L_i$ and an SLO threshold $\tau_i = \alpha \times T_{\mathrm{iso}}(i)$, where $T_{\mathrm{iso}}(i)$\footnote{We measure $T_{\mathrm{iso}}$ by recording the end-to-end latency of each workflow with maximum concurrency set to one in vLLM.} is its isolated latency and $\alpha$ is the slack factor. We define Goodput $\mathcal{G}(t)$ as:
\begin{equation}
    \mathcal{G}(t) = \frac{1}{\Delta t} \sum_{i \in \mathcal{W}_{\Delta t}} \mathbb{I}\big(L_i \le \alpha \times T_{\mathrm{iso}}(i)\big).
    \label{eq:goodput}
\end{equation}

Figure~\ref{fig:goodput_collapse} evaluates $\mathcal{G}(t)$ at $\alpha=3$ on ILR-1 and ILR-4 (\S\ref{sec:workload-analysis}).
The evaluated baselines struggle to sustain high goodput at all arrival rates, highlighting the difficulty of completing concurrent workflows within their latency budgets. Together with the prefix rebuild overhead in \S\ref{sec:rebuild_overhead}, this limitation motivates a serving system that accounts for both KV memory pressure and timely workflow progress during scheduling. MARS demonstrates that substantially higher goodput is achievable under the same memory constraints.

\begin{figure}[t]
    \centering
    \includegraphics[width=0.92\linewidth]{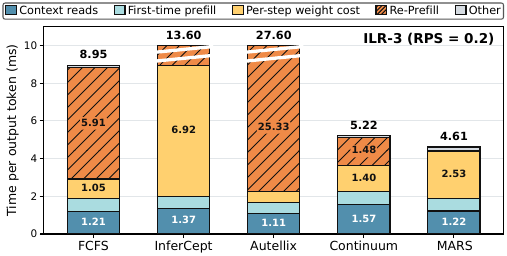}
    \vspace{-3mm}
    \caption{Breakdown of execution time per output token on ILR-3 at 0.2 RPS.}
    \label{fig:break}
    \vspace{-3mm}
\end{figure}

\subsection{Breakdown of Execution Time Per Token}
\label{sec:break_down}

Batching policies that emphasize token throughput often favor larger decode batches, which amortize weight access across more requests. Under GPU memory constraints, however, these savings can be outweighed by prefix rebuilds (or re-prefill) due to KV thrashing.
By contrast, smaller batches can alleviate memory contention, but they also weaken weight amortization. Choosing an appropriate batch size therefore requires balancing execution efficiency, hardware limitations, and KV thrashing.

Figure~\ref{fig:break} illustrates this trade-off on ILR-3 at 0.2 RPS by showing the breakdown of execution time per token.
FCFS, InferCept, and Autellix incur substantial re-prefill overhead.
Continuum also spends 1.48 ms per output token on re-prefill, comparable to its 1.40 ms amortized weight cost. These results highlight how repeated KV reconstruction can erode the benefits of weight amortization. MARS illustrates the potential for a better balance: despite a higher amortized weight cost, it nearly eliminates re-prefill and achieves a lower total time per output token than Continuum. Moreover, as shown in \S\ref{sec:goodput}, it also achieves higher workflow-level SLO-aware goodput. Together, these results demonstrate the opportunity to improve both execution efficiency and timely workflow completion under GPU memory constraints. We present the design that enables these improvements in the following sections.

\section{Overview}
\label{sec:arch_overview}

To achieve high workflow-level goodput under memory constraints, we design \sys, a memory-centric serving system that tames KV cache thrashing while accelerating per-turn completion. Our design addresses a fundamental \emph{granularity mismatch} in existing throughput-oriented engines: while reusable KV state and latency budgets belong to a multi-turn workflow, the engine admits and schedules work at the granularity of isolated inference calls. To maximize token throughput, existing engines greedily admit waiting calls into large batches whenever slots open. Consequently, optimizing in-engine call scheduling alone cannot ensure high goodput. Oblivious to live GPU memory pressure and workflow lifecycles, large-batch execution continuously admits incoming calls that evict the reusable KV states of unfinished workflows while slowing decode turns.

To break this cycle, \sys{} regulates the active KV working set to execute smaller, right-sized batches, shifting execution from evict-and-recover to admit-and-finish. Specifically, \sys decouples \emph{workflow-level working set admission} outside the engine from \emph{call-level execution and KV reclamation} inside the engine. External admission regulates which workflows enter the engine based on hardware capacity and initial prefix demand so workflows do not overfill GPU memory and trigger thrashing. Meanwhile, as admitted workflows continue to expand their contexts across turns, the in-engine scheduler manages post-admission memory pressure, prioritizes turn completion, and feeds live KV availability back to the admission controller.

\begin{figure}[t]
  \centering
  \includegraphics[width=\linewidth]{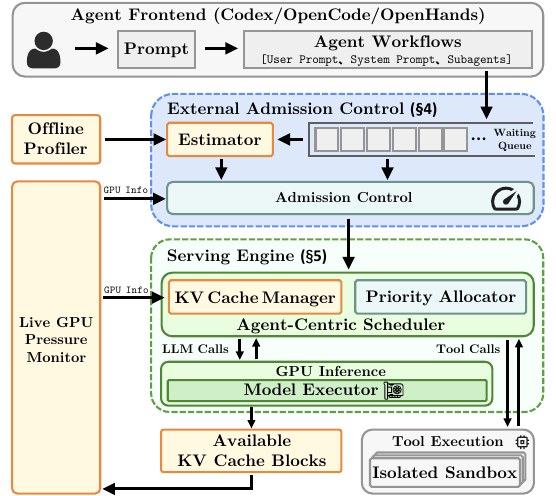}
  \vspace{-6mm}
  \caption{\sys architecture overview.}
  \label{fig:arch}
  \vspace{-3mm}
\end{figure}

Figure~\ref{fig:arch} presents the architecture of \sys, which adds an \emph{External Admission Controller} (\S\ref{sec:workset_admission}) and an \emph{Agent-Centric Scheduler} (\S\ref{sec:workset_scheduling}) to the underlying serving engine. Agent frontends retain control over tool selection, child-session creation, and workflow termination. Newly arrived root workflows wait in an external queue $Q_{\mathrm{wait}}$ until admitted into the active working set $\mathcal{A}$. Once admitted, a workflow remains in $\mathcal{A}$ across all inference turns and tool phases until completion, while its call readiness and KV residency evolve dynamically inside the engine.

The two components operate in a closed feedback loop. Outside the engine, the \emph{External Admission Controller} regulates both the concurrency window $W$ and the composition of $\mathcal{A}$ using offline hardware profiles, initial prefix-footprint estimates, and live free KV block counts, favoring smaller footprints under contention and draining large-prefix backlogs within live KV availability. Inside the engine, the \emph{Agent-Centric Scheduler} arbitrates ready calls within $\mathcal{A}$ by combining size- and service-aware priority scheduling with dynamic prefill chunk resizing and reverse-priority KV preemption. As calls allocate or release KV blocks, updated memory telemetry guides subsequent admissions.

\section{Working Set Admission Control}
\label{sec:workset_admission}

Agent workflows enter a serving replica with heterogeneous initial prefixes and continue to expand their contexts across subsequent turns. Consequently, regulating concurrency via a scalar workflow-count window, e.g., CONCUR~\cite{chen2026concur}, is insufficient: equal workflow counts can impose vastly different memory loads, while initial prefix sizes alone cannot capture post-admission growth. To prevent oversubscription, \sys regulates both the \emph{size} and the \emph{composition} of the active working set $\mathcal{A}$ by combining initial footprint estimation with live KV feedback.

\subsection{Live KV Capacity and Demand Estimation}
\label{sec:est}

\sys applies admission at the \emph{root workflow} level, which corresponds to an externally submitted task and may spawn child sessions during execution. Once admitted from $Q_{\mathrm{wait}}$ into $\mathcal{A}$, a root retains its admission across all inference turns and tool phases until termination, with its child sessions consuming no additional root slots in $\mathcal{A}$.

Consistent with serving engine allocators, \sys measures both memory supply and workflow demand in KV blocks.

At each admission opportunity, the controller reads the allocator's current allocatable block count $F$, including unreferenced blocks that may retain reusable prefix state.
It deducts a guard margin $B_{\mathrm{guard}}$ to account for allocation changes before the next update. For each waiting root $s$, the controller estimates its initial KV footprint $b_s$ from its input prefix length, using offline agent traces to configure $B_{\mathrm{guard}}$ and the threshold for large initial prefixes.

\subsection{Adaptive Workflow Admission}
\label{sec:req_admission}

As future model outputs and tool returns are unknown at admission time, \sys dynamically adjusts the target root concurrency window $W$ via a KV utilization feedback loop. The controller reduces $W$ multiplicatively when KV utilization exceeds a high watermark and increases $W$ additively when pressure subsides below a low watermark. Reducing $W$ throttles new root admissions without preempting roots already in $\mathcal{A}$.

An admission step is triggered upon workflow arrival, workflow termination, or periodic timer ticks while $Q_{\mathrm{wait}}$ is nonempty. Given the current window $W$ and free block count $F$, the controller computes the available root slots and KV block budget:
\[
    n = \max(0, W - |\mathcal{A}|),
    \qquad
    B = \max(0, F - B_{\mathrm{guard}}).
\]
When $n=0$, no new roots are admitted. Otherwise, the controller orders candidates in $Q_{\mathrm{wait}}$ based on memory pressure: under low pressure, it preserves FIFO arrival order; under contention, it prioritizes workflows with smaller initial footprints $b_s$ to limit the initial KV demand introduced by new roots. Scanning candidates in this order, the controller admits a workflow $s$ if $n > 0$ and $b_s \le B$, decrementing $n$ by $1$ and $B$ by $b_s$. Candidates that exceed the remaining budget $B$ are skipped for the current step to avoid head-of-line blocking, remaining in $Q_{\mathrm{wait}}$ for future opportunities.

\subsection{Working Set Construction and Lifecycle Semantics}
\label{sec:workset_construction}

By combining the adaptive window $W$ with the block budget $B$ and footprint estimates $b_s$, \sys moves beyond scalar concurrency control ~\cite{chen2026concur} and static peak-token admission~\cite{gong2025past}. These two control signals play orthogonal, complementary roles in shaping the active working set $\mathcal{A}$. Specifically, matching $b_s$ against $B$ limits the estimated initial demand introduced in each
admission step. Even when $W$ permits additional concurrency, the block-fit check prevents a burst of workflows with large initial prefixes from exhausting free KV blocks $F$ upon entry. Conversely, adapting $W$ via live KV telemetry reduces \emph{post-admission growth oversubscription}. When already-admitted workflows accumulate large tool outputs or spawn child sessions across turns, the resulting rise in KV utilization shrinks $W$, halting further root entries before the replica falls into thrashing.

Crucially, \sys preserves admission at the root-workflow level: subsequent turns and dynamically created child sessions inherit their parent root's admission and bypass $Q_{\mathrm{wait}}$. Requiring these calls to acquire additional root slots could block the progress of roots already in $A$, particularly when parents await child results. Inherited admission makes these calls eligible for in-engine scheduling; their execution and KV residency remain subject to the engine's scheduling and allocation checks. Across tool phases, reusable prefixes are retained opportunistically through the engine's prefix cache (\S\ref{sec:kv_block_manage}).

Finally, working set construction adapts smoothly across workload regimes. Under light load, ample $W$ and $B$ preserve arrival-order fairness. Under memory contention, prioritizing smaller $b_s$ admits roots with smaller initial KV footprints into $\mathcal{A}$, accelerating turn completions and reclaiming KV blocks for subsequent admissions. When this short-first preference eventually leaves $Q_{\mathrm{wait}}$ populated solely by deferred large-prefix workflows, \sys transitions to the KV-budgeted draining regime (\S\ref{sec:long_drain}) to make progress on the deferred tail while limiting additional memory pressure.

\section{Scheduling the Working Set}
\label{sec:workset_scheduling}

While external admission regulates which root workflows enter the active set $\mathcal{A}$, ready inference calls from admitted roots and their dynamically spawned child sessions still contend for finite GPU compute and KV blocks as their contexts grow. Uncoordinated execution within $\mathcal{A}$ can still stall multi-turn progress if long prefills monopolize the GPU or bursty allocations trigger cascading evictions.

\begin{figure}[t]
  \centering
  \includegraphics[width=\linewidth]{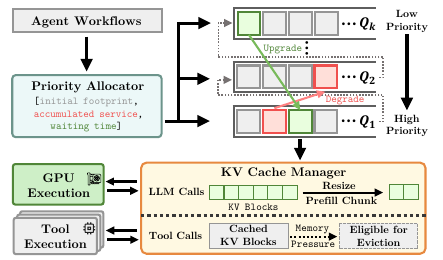}
  \vspace{-7mm}
  \caption{Detailed Agent-Centric Scheduler of \sys.}
  \label{fig:mlfq}
  \vspace{-3mm}
\end{figure}

To sustain rapid turn completions inside $\mathcal{A}$, \sys couples feedback-driven call prioritization with elastic, priority-aware KV management, as shown in Figure~\ref{fig:mlfq}. Ready inference calls are organized into multi-level priority queues $Q_1, \dots, Q_K$ ($Q_1$ highest). During batch construction, if free KV capacity cannot accommodate a proposed prefill, the scheduler elastically shrinks the prefill chunk to avoid evicting existing states. When reclamation becomes unavoidable, it preempts unfinished calls in reverse priority order ($Q_K \to Q_1$). Finally, when short-first admission leaves only large-prefix workflows in $Q_{\mathrm{wait}}$, \sys coordinates in-engine execution with KV-budgeted tail draining.

\subsection{Multi-Level Priority Inference}
\label{sec:priority}

\sys assigns and updates per-call priorities using three runtime signals: a call's initial KV footprint, its accumulated prefill service within the current turn, and its waiting time. Whereas root admission persists across the entire workflow lifecycle, scheduling state is instantiated independently for each inference call and discarded upon turn completion.

When an inference call arrives at the engine, the priority allocator maps its required prefill length and the configured token quantum to an initial queue, placing smaller prefills closer to $Q_1$ so lightweight turns can complete and enter tool execution quickly. As a call executes across chunked prefill iterations, the scheduler compares its remaining prefill length against its consumed prefill budget, using the lower of the two mapped priorities as its base level. Whenever a call's scheduled prefill tokens reach the quantum and prefill work remains, its priority is demoted by one level. This demotion prevents long prefills from monopolizing GPU compute or rapidly exhausting free KV blocks at the expense of co-running calls.

Once a call finishes prefilling and enters decoding, its priority is immediately promoted to $Q_1$, and subsequent decode tokens do not consume the prefill quantum. Prioritizing active decodes directly accelerates per-turn completion and sustains decode efficiency (Figure~\ref{fig:break}), allowing sessions to invoke their next tool or finish and release KV capacity. To mitigate prolonged waiting among demoted prefills, an aging mechanism boosts a waiting call's priority by one level for every 30 seconds since its last service, up to $Q_1$. Within each queue $Q_k$, calls are dispatched in FIFO order.

If a running call is preempted under memory pressure, \sys resets the call's scheduling feedback and requeues it at the front of $Q_1$ with reset scheduling feedback. When a call finishes decoding and leaves the engine for tool execution, its call-level priority state is discarded; when the workflow returns for its next turn, the new call initializes its priority freshly from its updated context demand while retaining engine access through its parent root's admission in $\mathcal{A}$.

\subsection{Elastic Prefill Resizing and KV Reclamation}
\label{sec:kv_block_manage}

During each scheduling iteration, \sys constructs the execution batch from $Q_1$ to $Q_K$ while keeping KV capacity checks. When a scheduled prefill chunk exceeds the currently available KV blocks, traditional engines immediately trigger eviction or offloading. In contrast, \sys first attempts \emph{elastic prefill resizing}: it iteratively shrinks the call's proposed prefill chunk size to fit the remaining free blocks, using a single KV block as the lower bound. By trading a smaller single-iteration chunk size for eviction avoidance, elastic resizing allows prefills to advance incrementally within tight memory headroom to avoid displacing reusable states. If a waiting call cannot fit even at the single-block lower bound, it remains queued and the scheduler stops adding new waiting calls to the current batch.

For an already-running call whose minimum prefill chunk or next decode step still cannot obtain a KV block, \sys performs priority-aware KV reclamation. It selects unfinished calls holding active KV references in reverse priority order, from $Q_K$ to $Q_1$, excluding calls committed to the current batch, and stops immediately once enough blocks are freed to satisfy the pending allocation. Preempted calls release their active KV block references and return to the waiting queues, while their root workflows remain admitted in $\mathcal{A}$. This reverse-priority preemption preferentially preserves active decodes and higher-priority prefills while reclaiming capacity from lower-priority calls.

Across external tool phases, \sys leverages the engine's block-level prefix cache for opportunistic state reuse. Upon completing a turn, a call releases its active block references, converting full KV blocks into cached, evictable prefix blocks without retaining the completed call's priority. As external admission (\S\ref{sec:workset_admission}) bounds the aggregate working set in $\mathcal{A}$, these cached blocks rarely suffer premature displacement during tool intervals, enabling returning calls to hit the prefix cache and avoid costly prefix rebuilds.

\subsection{KV-Aware Draining of Deferred Long Workflows}
\label{sec:long_drain}

Prioritizing smaller initial footprints under contention (\S\ref{sec:req_admission}) can eventually leave the external queue $Q_{\mathrm{wait}}$ dominated by deferred root workflows with large initial prefixes. Once the backlog enters this tail regime, call-level prioritization offers diminishing opportunities for early completions, whereas admitting multiple large-prefix workflows concurrently would immediately recreate KV oversubscription and in-engine thrashing. \sys therefore coordinates tail completion through \emph{KV-budgeted draining}, bounding the concurrent exposure of large-prefix roots.

The admission controller identifies large-prefix root workflows using the configured threshold and activates draining once only such roots remain in $Q_{\mathrm{wait}}$. In this mode, the controller reverts to arrival-order scanning over $Q_{\mathrm{wait}}$ and admits a subset of workflows whose estimated initial footprints $b_s$ fit within the live block budget $B = \max(0, F - B_{\mathrm{guard}})$. Candidates exceeding the remaining budget $B$ are skipped for the current step, preventing head-of-line blocking while preserving arrival order as the primary discipline.

To prevent an abrupt surge in KV demand when earlier workflows are still executing in $\mathcal{A}$, \sys \emph{soft-starts} the draining phase by admitting at most one large-prefix workflow per admission opportunity while prior workflows remain active. Furthermore, if no active workflows remain in $\mathcal{A}$ yet the oldest waiting workflow's initial footprint $b_s$ exceeds the conservative guard budget $B$, \sys admits that oldest workflow alone so the controller never deadlocks on $B_{\mathrm{guard}}$. As admitted tail workflows advance through turns using the priority scheduling and elastic KV management of \S\ref{sec:priority}--\ref{sec:kv_block_manage} and eventually terminate, freed KV blocks replenish $B$ and trigger subsequent draining admissions.

\section{Implementation}
\label{sec:implementation}

\noindent \textbf{Integration with vLLM.}
We implement \sys in approximately 5,000 lines of Python on top of vLLM V1~\cite{kwon2023vllm}. The external controller maintains $Q_{wait}$, $\mathcal{A}$, and $W$, and polls KV statistics exposed by the engine. We extend the OpenAI-compatible request schema with metadata, including a stable root \texttt{job\_id} for attribution, and propagate these fields through \texttt{SamplingParams.extra\_args} to the V1 request objects. Scheduler extensions implement priority feedback and priority-aware execution and preemption.

\noindent \textbf{Integration with Agent Frontends.}
We integrate \sys with OpenHands~\cite{wang2025openhands}, OpenCode~\cite{opencode2025}, and Codex CLI~\cite{openaiCodexPrompting2026}. Each admitted root runs in a dedicated worker process. Frontend adapters collect LLM and tool events and attach the root's \texttt{job\_id} to model requests, including calls from child sessions. OpenHands uses instrumented runtime boundaries; OpenCode and Codex connect through a forwarding proxy that handles API-schema compatibility where required. The frontends retain control over tool execution and delegation, while \sys applies root-level admission and call-level scheduling without requiring a predeclared agent graph.

\section{Evaluation}
\label{sec: eval}

\subsection{Workload Analysis}
\label{sec:workload-analysis}

We construct a hybrid workload pool from five agentic and long-context benchmarks: SWE-bench~\cite{jimenez2024swebench}, GitTaskBench~\cite{ni2025gittaskbench}, Terminal-Bench~\cite{merrill2026terminalbench}, RepoBench~\cite{liu2024repobench}, and $\infty$Bench~\cite{zhang2024inftybench}. From this pool, we sample four workloads of 400 root workflows each, denoted \textsc{ILR-1} through \textsc{ILR-4} by their \textsc{Input Length Regimes}. To evaluate models with smaller context limits (GPT-OSS-120B) and tighter VRAM budgets, we similarly sample three Short-ILR workloads (\textsc{S-ILR1}--\textsc{S-ILR3}) with a lower context-length cap.

\begin{figure}[t!]
    \centering
    \includegraphics[width=\linewidth]{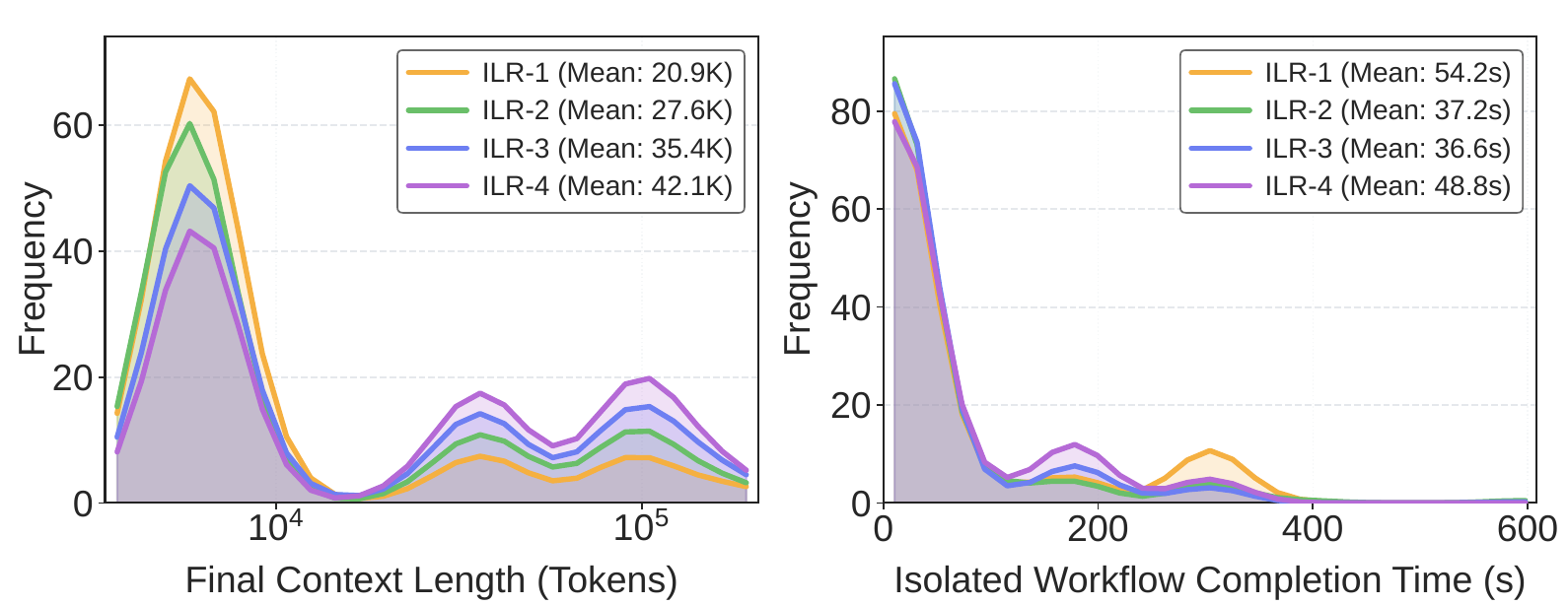}
    \vspace{-6mm}
    \caption{Workload characterization across four input-length regimes. Left: final workflow context length. Right: isolated workflow completion time $T_{\mathrm{iso}}$.}
    \label{fig:workload-analysis}
    \vspace{-3mm}
\end{figure}

Figure~\ref{fig:workload-analysis} plots the distributions of final context length and isolated completion time ($T_{\mathrm{iso}}$) across \textsc{ILR-1}--\textsc{ILR-4}. While mean final context length grows monotonically from 20.9K to 42.1K tokens, mean isolated completion times remain comparable (54.2\,s, 37.2\,s, 36.6\,s, and 48.8\,s) with broadly overlapping distributions. These regimes thus isolate the impact of escalating KV working set pressure from intrinsic execution duration.

\begin{figure*}[!t]
    \centering
    \includegraphics[width=\textwidth]{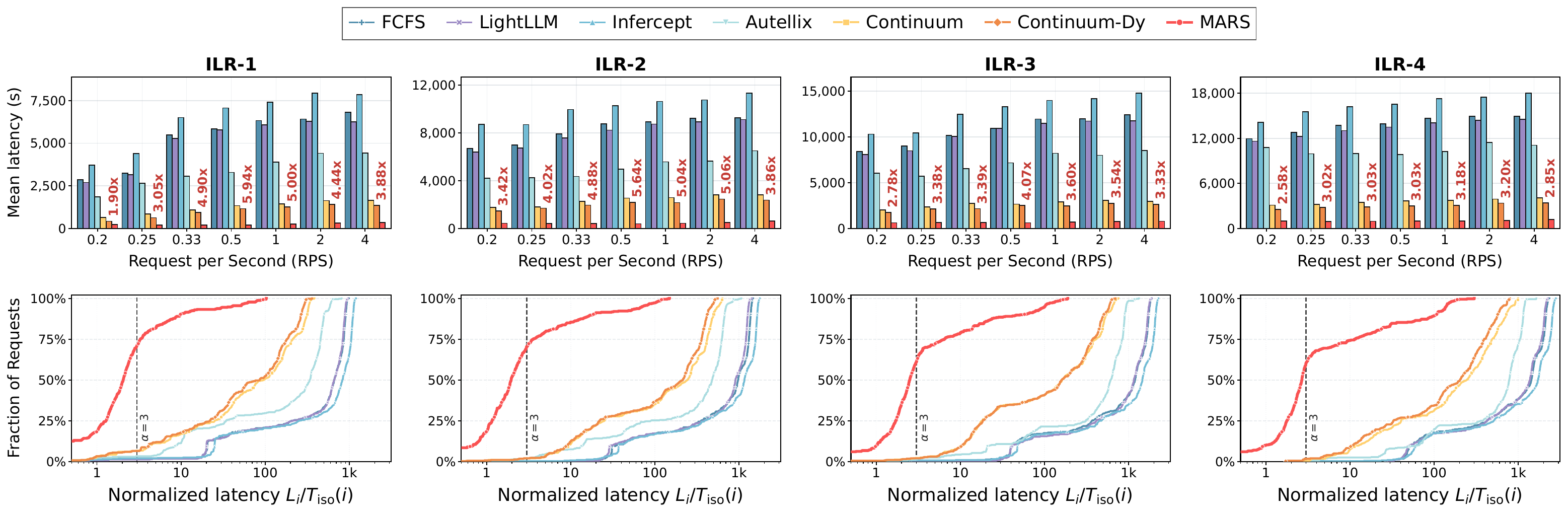}
    \vspace{-8mm}

    \caption{The first row reports mean end-to-end latencies for Qwen3-Coder-30B on the H200 testbed across \textsc{ILR-1} to \textsc{ILR-4}. Red text denotes the speedup of \sys over the fastest baseline. The second row plots the CDF of normalized workflow latency $L_i/T_{\mathrm{iso}}(i)$. The CDF at $\alpha$ equals the fraction of requests satisfying the per-task SLO. }
    \label{fig:main_h200}
    \vspace{-3mm}
\end{figure*}

\begin{figure*}[htbp]
    \centering
    \includegraphics[width=0.96\linewidth]{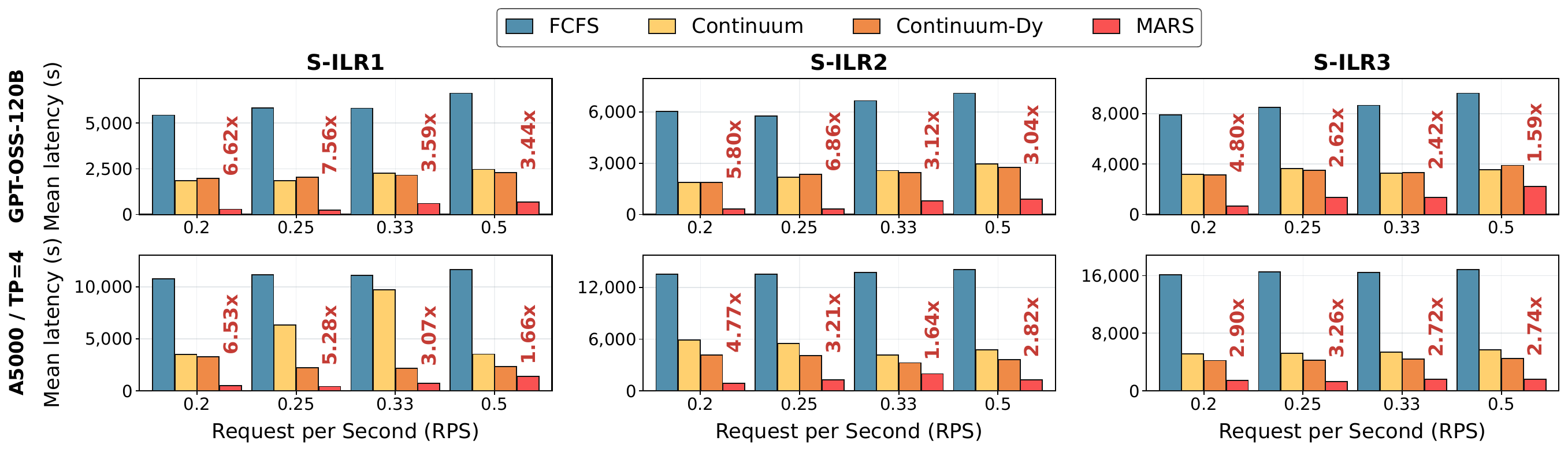}
    \vspace{-5mm}
    \caption{\textbf{Generality across models and GPU configurations.} Mean end-to-end latency across three input-length regimes (S-ILR1-S-ILR3). The top row evaluates GPT-OSS-120B on a single H100, while the bottom row evaluates Qwen3-Coder-30B using tensor parallelism across four A5000 GPUs (TP=4). }
    \label{fig:model_hardware}
    \vspace{-3mm}
\end{figure*}

\subsection{Experimental Setup}
\label{sec:exp-setup}

\noindent \textbf{Models \& Testbed.}
We evaluate \sys using Qwen3-Coder-30B-A3B-Instruct (262K context limit)~\citep{qwen3} and GPT-OSS-120B (131K context limit)~\citep{openai_gpt_oss_2025}. Single-GPU experiments ($\mathrm{TP}=1$) run on either an NVIDIA H200 NVL (144\,GiB VRAM, AMD EPYC 9355) or an NVIDIA H100 NVL (96\,GiB VRAM, AMD EPYC 9554). Multi-GPU experiments ($\mathrm{TP}=4$) run across four NVIDIA RTX A5000 GPUs (24\,GiB each, AMD EPYC 7352) within a single NUMA domain. The serving engine and tool workers are pinned to disjoint physical CPU cores. We evaluate both an online task-driver testbed over our modified vLLM backend and full-stack deployments with OpenHands, OpenCode, and Codex CLI frontends.

\noindent \textbf{Metrics.}
Root workflows arrive according to a Poisson process across varying request rates (RPS). We report \textbf{end-to-end workflow latency} (spanning all inference turns and tool phases from submission to termination) and \textbf{per-task SLO attainment} ($L_i \le \alpha \times T_{\mathrm{iso}}(i)$, §\ref{sec:goodput}) across all submitted workflows.

\noindent \textbf{Baselines.}
To provide a fair comparison, all baselines are implemented in our modified vLLM V1 runtime, sharing identical tool-execution stacks, maximum batch sizes, prompts, arrival timestamps, and CPU/GPU partitioning:
\begin{itemize}[leftmargin=*, nosep, topsep=2pt]
    \item \textbf{FCFS:} vLLM's default first-come-first-served scheduler, representing throughput-centric continuous batching.
    \item \textbf{LightLLM\footnote{https://github.com/ModelTC/lightllm}}: Admits waiting requests only when their estimated peak token demand fits available KV capacity. We also evaluate unmodified LightLLM with OpenHands.
    \item \textbf{Autellix}~\cite{autellix}: An agent-aware system that prioritizes agent calls using Program-Level Attained Service (PLAS) to favor shorter workflows.
    \item \textbf{InferCept\footnote{https://github.com/WukLab/InferCept}}~\cite{infercept}: Selectively preserves, swaps (implemented via LMCache~\cite{liu2025lmcache}), or evicts KV states across tool intervals using execution-time heuristics.
    \item \textbf{Continuum\footnote{https://github.com/Hanchenli/vllm-continuum}}~\cite{continuum}: Applies phase-aware KV cache time-to-live (TTL) retention at tool boundaries. We evaluate both static Continuum in its source code and augmented dynamic heuristic-driven \textbf{Continuum-Dy}.
\end{itemize}

\begin{figure*}[t]
\centering
\includegraphics[width=\linewidth]{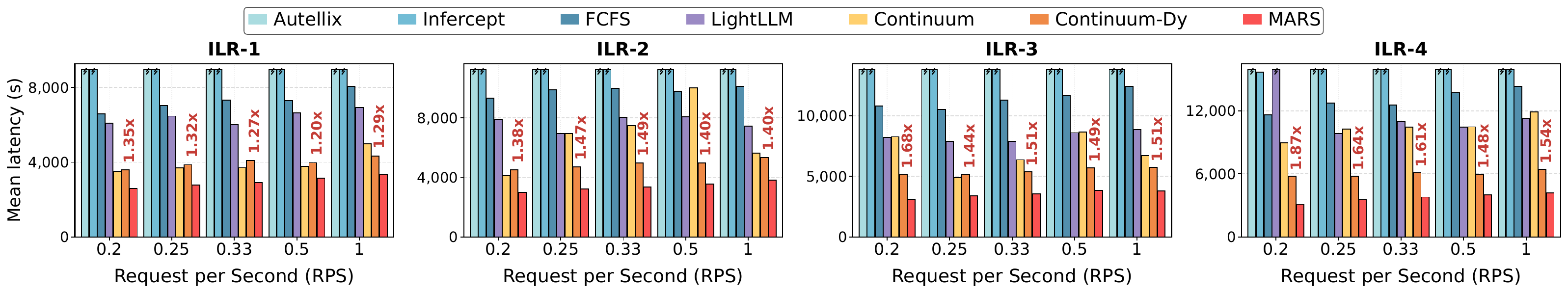}
\vspace{-8mm}
\caption{Mean end-to-end latency for Qwen3-Coder-30B on H200 in the full-stack OpenHands deployment across \textsc{ILR-1} to \textsc{ILR-4}. Red text denotes the speedup of \sys over the fastest baseline. Diagonal break marks (//) indicate bars whose measured mean latency exceeds the displayed y-axis range.}
\label{fig:Openhands}
\vspace{-3mm}
\end{figure*}

\begin{figure}[t]
\centering
\includegraphics[width=\linewidth]{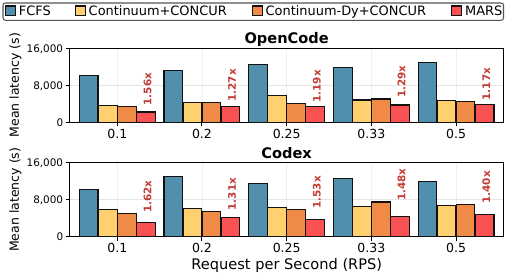}
\vspace{-8mm}
\caption{Mean end-to-end latency with OpenCode and Codex using Qwen3-Coder-30B under ILR-1 on H100.}
\label{fig:frontend}
\vspace{-3mm}
\end{figure}

\subsection{End-to-End Serving Performance}
\label{sec:eval-e2e}

\noindent \textbf{Mean End-to-End Latency.}
We first evaluate the mean end-to-end performance in Figure~\ref{fig:main_h200}. The top row shows the result of Qwen3-Coder-30B on the H200 testbed across four input-length regimes and seven arrival rates. \sys achieves the lowest mean latency across all 28 operating points, delivering $1.90\times$--$5.94\times$ speedups over the fastest baseline at each point. By bounding the active working set ($\mathcal{A}$) and avoiding premature prefix eviction (§\ref{sec:workset_admission}--\ref{sec:workset_scheduling}), \sys eliminates the rebuild cascades that stall baseline workflows as arrival rate or context length increases.

\noindent \textbf{Per-Task SLO Attainment.}
The bottom row of Figure~\ref{fig:main_h200} plots the CDF of normalized workflow latency $L_i/T_{\mathrm{iso}}(i)$ at 0.5\,RPS, where the curve value at $\alpha=3$ (vertical dashed line) gives the fraction of workflows meeting their per-task SLO (§\ref{sec:goodput}). At $\alpha=3$, \sys{} satisfies the SLO for 71\% of workflows versus 7\% for the strongest baseline in \textsc{ILR-1}, and 60\% versus 2\% in \textsc{ILR-4}, with similar distributional shifts in \textsc{ILR-2} and \textsc{ILR-3}. \sys eliminates in-engine KV thrashing and ensures timely end-to-end workflow completions across the input configurations.

\subsection{Generality Across Models and Hardware}
\label{sec:eval-generality}

We next vary the backend model and execution configuration to test \sys's generality.

\noindent \textbf{H100 + GPT-OSS-120B.}
The top row of Figure~\ref{fig:model_hardware} evaluates the larger GPT-OSS-120B model on H100 (96\,GiB VRAM) across \textsc{S-ILR1}--\textsc{S-ILR3}. \sys outperforms the fastest baseline by $1.59\times$--$7.56\times$ across all points. As a 120B model leaves a smaller VRAM budget for the KV pool and incurs higher compute cost per rebuilt token, premature KV eviction and swapping carry steeper penalties, amplifying the benefit of \sys's memory-centric working set control.

\noindent \textbf{A5000 + Qwen3-Coder-30B (TP=4).}
The bottom row of Figure~\ref{fig:model_hardware} deploys Qwen3-Coder-30B across four 24\,GiB A5000 GPUs using tensor parallelism ($\mathrm{TP}=4$). \sys achieves $1.64\times$--$6.53\times$ speedups over the fastest baseline. While tensor parallelism expands available VRAM across devices, it introduces inter-GPU communication and still exposes the replica to KV thrashing when uncoordinated requests overflow the engine. By regulating admission and KV residency above the TP execution layer, \sys preserves substantial latency reductions on multi-GPU hardware.

\subsection{Integration with Agent Frontends}
\label{sec:eval-openhands}

Finally, we integrate \sys as the backend of mainstream agent frontends to evaluate the full agent loop performance.

\noindent \textbf{OpenHands.}
Figure~\ref{fig:Openhands} reports full-stack performance when serving OpenHands with Qwen3-Coder-30B on H200, including its native control loop, RPCs, and containerized tool sandboxes. Across all 20 operating points in \textsc{ILR-1}--\textsc{ILR-4}, \sys achieves the lowest mean completion time, yielding $1.20\times$--$1.87\times$ speedups over the fastest baseline.

\noindent \textbf{OpenCode and Codex.}
Figure~\ref{fig:frontend} evaluates OpenCode and Codex on H100 under \textsc{ILR-1}, where both frontends dynamically invoke tools and spawn child sessions at runtime without exposing a predeclared agent graph. Compared against FCFS, Continuum + CONCUR, and Continuum-Dy + CONCUR, \sys{} reduces mean completion time by $1.17\times$--$1.56\times$ on OpenCode and $1.31\times$--$1.62\times$ on Codex.

\noindent \textbf{Understanding the Full-Stack Gains.}
The smaller speedups with OpenHands, OpenCode, and Codex are consistent with serving optimizations affecting only part of the end-to-end execution path. Full-stack integrations retain each framework's native orchestration, where frontend processing, RPCs, sandbox management, and tool execution occupy a sizable share of workflow latency and dilute backend reductions in admission delay, inference queuing, and KV recovery. The benefit of prefix-aware admission and service-based prioritization also depends on backlog composition. When long-running workflows with sequential LLM--tool dependencies dominate, reordering offers fewer opportunities to bring completions forward. Despite these full-stack bounds, \sys still achieves the lowest mean latency at every point across all frontends ($1.17\times$--$1.87\times$).

\begin{figure}[t]
\centering
\includegraphics[width=\linewidth]{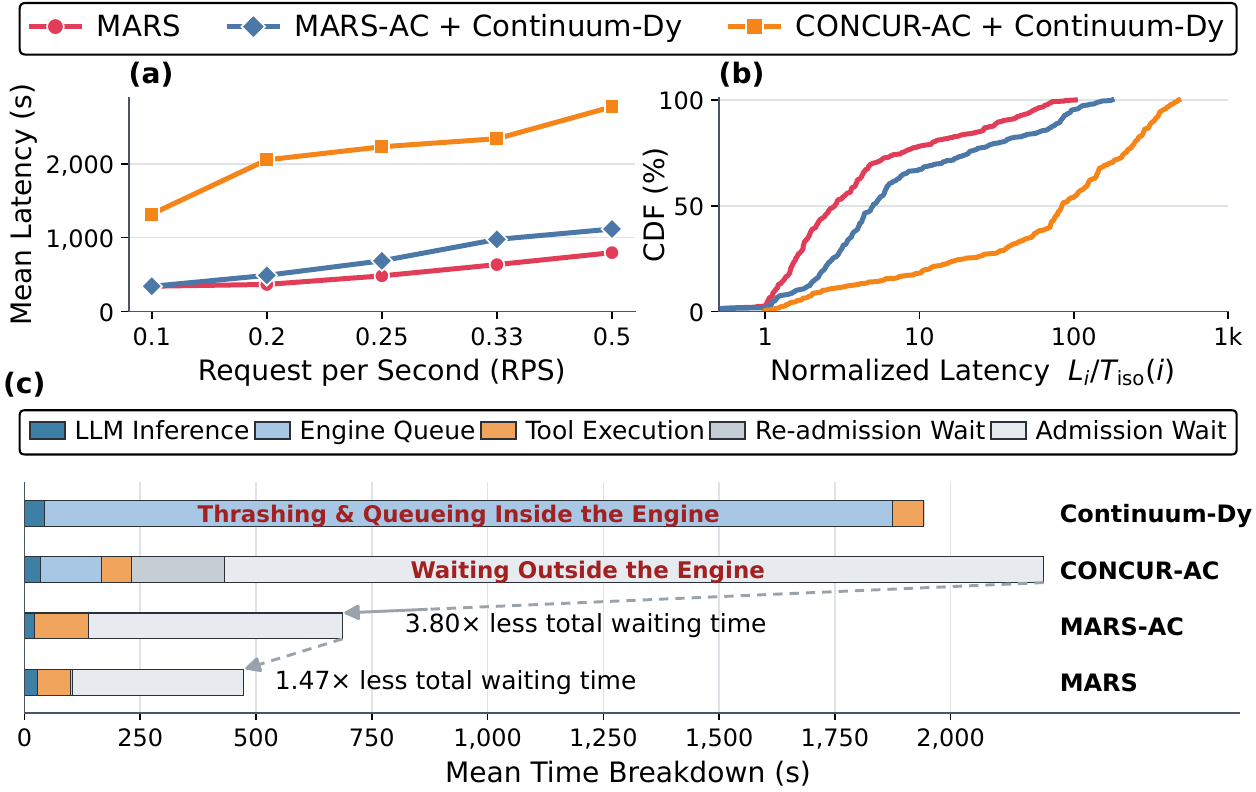}
\vspace{-8mm}
\caption{Ablation study on H100 with Qwen3-Coder-30B under ILR-1. (a): Mean end-to-end latency of full \sys, \sys -AC (Admission Controller) with Continuum-Dy, and CONCUR-AC with Continuum-Dy. (b)(c): CDF of normalized end-to-end latency and mean time breakdown at 0.25 requests/s.}
\label{fig:admission_ablation}
\vspace{-3mm}
\end{figure}

\subsection{Ablation Study}
\label{sec:eval_admission_ablation}

We now isolate the contributions of \sys's two core components: the external admission controller and the in-engine scheduler, on H100 using Qwen3-Coder-30B under ILR-1. We compare three variants under identical runtime settings: (1) \textbf{CONCUR-AC + Continuum-Dy} combines CONCUR's cache-feedback admission~\cite{chen2026concur} with Continuum-Dy, our strongest baseline. (2) \textbf{\sys{}-AC + Continuum-Dy} retains Continuum-Dy's scheduling and KV management but replaces CONCUR-AC with \sys's working set admission controller; and (3) \textbf{Full \sys{}}.

Figure~\ref{fig:admission_ablation}(a) shows that \sys's admission controller jointly controls active set breadth and composition under heterogeneous KV demand and outperforms the scalar CONCUR admission controller by 2.4$\times$--4.2$\times$. Full \sys performs similarly at low load, but provides a further $1.33\times$--$1.54\times$ speedup over \sys-AC + Continuum-Dy from $0.2$ to $0.5$ requests/s as contention increases, and outperforms CONCUR-AC + Continuum-Dy by $3.47\times$--$5.59\times$.

Figure~\ref{fig:admission_ablation}(b) shows a consistent leftward shift in normalized latency across the workflow distribution, indicating that the improvement extends beyond the mean latency. Overall, working set admission provides the primary gain, while coordinated scheduling and KV management become increasingly important under higher contention.

Figure~\ref{fig:admission_ablation}(c) shows where the latency gains come from.  Without external admission control (\textbf{Continuum-Dy}), requests accumulate inside the engine, where KV thrashing and queueing impede progress under memory pressure. CONCUR-AC limits this exposure and shifts much of the delay to waiting outside the engine, but its scalar window counts workflows, not their KV demand. \sys-AC adds live KV capacity and per-workflow prefix estimates to an adaptive window, regulating both the size and the composition of the admitted set; mean total waiting falls from CONCUR-AC's 2,101\,s to 553\,s ($3.80\times$). Full \sys further adds coordinated execution and KV reclamation, which curb disruptive allocations and repeated prefix reconstruction inside the engine, bringing mean waiting to 375\,s (another $1.47\times$).

\section{Discussion}

\noindent \textbf{Objective, Fairness, and Starvation.}
\sys optimizes mean end-to-end latency and SLO-aware goodput. Short-first admission under contention defers workflows with large initial footprints, and KV-budgeted tail draining together with in-engine priority aging mitigate starvation. Tail draining addresses the deferred remainder of a finite backlog. However, tenant-level resource sharing is outside this objective. Extending \sys with a fairness policy such as dominant resource fairness~\cite{ghodsi2011dominant} would require accounting for demand that shifts as workflows alternate between inference and tools and spawn child sessions.

\noindent \textbf{Beyond a Single Serving Replica.}
\sys currently manages one node-local replica, spanning one or more GPUs through tensor parallelism. Scaling to multi-replica clusters requires coordinating workflow routing, admission, and load rebalancing under per-replica KV pressure. While session affinity preserves KV locality, it can skew memory load across replicas; conversely, live migration as in Llumnix~\cite{llumnix} relieves imbalance at the cost of KV transfer or rebuilds. Deciding when to retain, migrate, or reconstruct state across tool intervals remains future work.

\noindent \textbf{Dependency-Aware Scheduling.}
\sys schedules each ready call as an independent unit of work under their root's admission, supporting dynamically created subagents without a predeclared graph. In tree-search~\cite{yao2024tot} and multi-agent workloads, the calls from one root differ in agent graph topology. When a frontend exposes such dependencies, \sys's priority allocator could weight calls by downstream dependents and its reclamation could prefer victims off the critical path, extending admit-and-finish from individual roots to the structure within them.

\section{Related Work}

\noindent \textbf{Throughput-Centric Inference Systems.}
Modern serving engines maximize token throughput via iteration-level scheduling in Orca~\cite{yu2022orca}, PagedAttention in vLLM, chunked or split prefill--decode batching in Sarathi-Serve~\cite{agrawal2024sarathi} and DeepSpeed-FastGen~\cite{ds_fastgen}, and fault-tolerant instance scheduling in SpotServe~\cite{miao2024spotserve}. \sys builds atop these execution primitives, regulating the workflow working set to sustain both high token throughput and stable end-to-end workflow progress.

\noindent \textbf{Agentic Inference Systems.}
Recent systems optimize multi-call state reuse and scheduling from several angles. SGLang~\cite{zheng2024sglang}, PromptCache~\cite{gim2023promptcache}, and CacheGen~\cite{liu2024cachegen} accelerate KV reuse and transfer, while Autellix~\cite{autellix}, InferCept~\cite{infercept}, Continuum~\cite{continuum}, CONCUR~\cite{chen2026concur}, and JITServe~\cite{zhang2026jitserve} improve dynamic agent serving through attained-service scheduling, tool-interval KV management, concurrency windows, or SLO-aware batching. A separate class of systems, including Ayo~\cite{ayo}, ALTO~\cite{alto}, Parrot~\cite{lin2024parrot}, and TokenCake~\cite{bian2025tokencake}, optimizes multi-agent workflows by exploiting predeclared execution graphs or explicit cross-call dependencies. As these graph-based approaches require upfront workflow topologies that dynamic agent frontends do not expose at runtime, they fall outside the scope of \sys.
\sys targets workflows whose execution structure evolves at runtime. Its policies remain agnostic to workflow topology and coordinate working set admission, call prioritization, and KV reclamation through live memory feedback.

\noindent \textbf{LLM-OS Abstractions.}
AIOS~\cite{aios} and AgentOS~\cite{agentos} provide OS-style kernels for agent orchestration and resource isolation, while MemGPT~\cite{memgpt} manages virtual context across memory tiers. These frontend and context abstractions are orthogonal to \sys{}, which prevents KV thrashing inside the serving backend while preserving the frontend's native control loop.

\section{Conclusion}
\label{sec:conclusion}

As autonomous LLM agents are increasingly deployed locally, their expanding contexts easily exhaust limited GPU memory, triggering KV cache thrashing and degrading workflow-level goodput. We present \sys, a memory-centric serving system that reconciles low KV thrashing with high SLO attainment by executing smaller, memory-bounded working sets. \sys shifts execution from evict-and-recover to admit-and-finish by coupling an external admission controller, guided by initial prefix footprints and live KV availability, with an in-engine agent-centric scheduler for priority turn completion, elastic prefill resizing, and priority-aware KV reclamation. Built on vLLM, \sys reduces mean end-to-end latency by $1.59\times$--$7.56\times$ over the fastest baseline across diverse models and GPUs, and speeds up workflow completion in OpenHands, OpenCode, and Codex by $1.17\times$--$1.87\times$.

\section{Acknowledgments}
\label{sec:ack}
This work was supported in part by the National Science Foundation (NSF) (Grant No. 2625580) and the Army Research Office (ARO) (Grant No. W911NF-23-2-0224).

{\small
\bibliographystyle{mlsys2025}
\bibliography{references}

@inproceedings{infercept,
  author    = {Reyna Abhyankar and Zijian He and Vikranth Srivatsa and
               Hao Zhang and Yiying Zhang},
  title     = {{InferCept}: Efficient Intercept Support for Augmented Large
               Language Model Inference},
  booktitle = {Proceedings of the 41st International Conference on Machine Learning
               (ICML 2024)},
  series    = {Proceedings of Machine Learning Research},
  volume    = {235},
  pages     = {81--95},
  year      = {2024}
}

@inproceedings{agrawal2024sarathi,
  author    = {Amey Agrawal and Nitin Kedia and Ashish Panwar and
               Jayashree Mohan and Nipun Kwatra and Bhargav Gulavani and
               Alexey Tumanov and Ramachandran Ramjee},
  title     = {Taming {Throughput-Latency} Tradeoff in {LLM} Inference with
               {Sarathi-Serve}},
  booktitle = {18th USENIX Symposium on Operating Systems Design and
               Implementation (OSDI 2024)},
  pages     = {117--134},
  year      = {2024}
}

@misc{chase2022langchain,
  author    = {Harrison Chase},
  title     = {{LangChain}},
  year      = {2022},
  url       = {https://github.com/langchain-ai/langchain}
}

@inproceedings{chen2026concur,
  author    = {Qiaoling Chen and Zhisheng Ye and Tian Tang and Peng Sun and
               Boyu Tian and Guoteng Wang and Shenggui Li and Yonggang Wen and
               Zhenhua Han and Tianwei Zhang},
  title     = {{CONCUR}: High-Throughput Agentic Batch Inference of {LLM} via
               Congestion-Based Concurrency Control},
  booktitle = {Forty-Third International Conference on Machine Learning
               (ICML 2026)},
  year      = {2026}
}

@inproceedings{erdogan2024tinyagent,
  author    = {Lutfi Eren Erdogan and Nicholas Lee and Siddharth Jha and
               Sehoon Kim and Ryan Tabrizi and Suhong Moon and
               Coleman Richard Charles Hooper and Gopala Anumanchipalli and
               Kurt Keutzer and Amir Gholami},
  title     = {{TinyAgent}: Function Calling at the Edge},
  booktitle = {Proceedings of the 2024 Conference on Empirical Methods in
               Natural Language Processing: System Demonstrations
               (EMNLP 2024)},
  pages     = {80--88},
  year      = {2024}
}

@inproceedings{ghodsi2011dominant,
  author    = {Ali Ghodsi and Matei Zaharia and Benjamin Hindman and
               Andy Konwinski and Scott Shenker and Ion Stoica},
  title     = {Dominant Resource Fairness: Fair Allocation of Multiple
               Resource Types},
  booktitle = {8th USENIX Symposium on Networked Systems Design and
               Implementation (NSDI 2011)},
  pages     = {323--336},
  year      = {2011}
}

@article{gim2023promptcache,
  author    = {Gim, In and Chen, Guojun and Lee, Seung-seob and
               Sarda, Nikhil and Khandelwal, Anurag and Zhong, Lin},
  title     = {Prompt Cache: Modular Attention Reuse for Low-Latency Inference},
  journal   = {Proceedings of Machine Learning and Systems (MLSys 2024)},
  volume    = {6},
  pages     = {325--338},
  year      = {2024}
}

@inproceedings{gong2025past,
  author    = {Ruihao Gong and Shihao Bai and Siyu Wu and Yunqian Fan and
               Zaijun Wang and Xiuhong Li and Hailong Yang and Xianglong Liu},
  title     = {{Past-Future} Scheduler for {LLM} Serving under {SLA} Guarantees},
  booktitle = {Proceedings of the 30th ACM International Conference on
               Architectural Support for Programming Languages and Operating
               Systems (ASPLOS 2025), Volume 2},
  pages     = {798--813},
  year      = {2025}
}

@article{ds_fastgen,
  author    = {Connor Holmes and Masahiro Tanaka and Michael Wyatt and
               Ammar Ahmad Awan and Jeff Rasley and Samyam Rajbhandari and
               Reza Yazdani Aminabadi and Heyang Qin and Arash Bakhtiari and
               Lev Kurilenko and Yuxiong He},
  title     = {{DeepSpeed-FastGen}: High-Throughput Text Generation for {LLMs}
               via {MII} and {DeepSpeed-Inference}},
  journal   = {arXiv preprint arXiv:2401.08671},
  year      = {2024}
}

@inproceedings{hong2024metagpt,
  author    = {Sirui Hong and Mingchen Zhuge and Jonathan Chen and Xiawu Zheng and
               Yuheng Cheng and Jinlin Wang and Ceyao Zhang and Zili Wang and
               Steven Ka Shing Yau and Zijuan Lin and Liyang Zhou and Chenyu Ran and
               Lingfeng Xiao and Chenglin Wu and J{\"u}rgen Schmidhuber},
  title     = {{MetaGPT}: Meta Programming for A Multi-Agent Collaborative Framework},
  booktitle = {The Twelfth International Conference on Learning Representations
               (ICLR 2024)},
  year      = {2024}
}

@inproceedings{jimenez2024swebench,
  author    = {Carlos E. Jimenez and John Yang and Alexander Wettig and
               Shunyu Yao and Kexin Pei and Ofir Press and
               Karthik R. Narasimhan},
  title     = {{SWE}-bench: Can Language Models Resolve Real-World {GitHub} Issues?},
  booktitle = {The Twelfth International Conference on Learning Representations
               (ICLR 2024)},
  year      = {2024}
}

@inproceedings{kwon2023vllm,
  author    = {Woosuk Kwon and Zhuohan Li and Siyuan Zhuang and Ying Sheng and
               Lianmin Zheng and Cody Hao Yu and Joseph E. Gonzalez and
               Hao Zhang and Ion Stoica},
  title     = {Efficient Memory Management for Large Language Model Serving
               with {PagedAttention}},
  booktitle = {Proceedings of the 29th ACM Symposium on Operating Systems
               Principles (SOSP 2023)},
  pages     = {611--626},
  year      = {2023}
}

@article{continuum,
  author    = {Hanchen Li and Runyuan He and Qiuyang Mang and Qizheng Zhang and
               Huanzhi Mao and Xiaokun Chen and Hangrui Zhou and Alvin Cheung and
               Joseph Gonzalez and Ion Stoica},
  title     = {Continuum: Efficient and Robust Multi-Turn {LLM} Agent Scheduling
               with {KV} Cache Time-to-Live},
  journal   = {arXiv preprint arXiv:2511.02230},
  year      = {2025}
}

@inproceedings{lin2024parrot,
  author    = {Chaofan Lin and Zhenhua Han and Chengruidong Zhang and
               Yuqing Yang and Fan Yang and Chen Chen and Lili Qiu},
  title     = {Parrot: Efficient Serving of {LLM-based} Applications with
               Semantic Variable},
  booktitle = {18th USENIX Symposium on Operating Systems Design and
               Implementation (OSDI 2024)},
  pages     = {929--945},
  year      = {2024}
}

@article{agentos,
  author    = {Rui Liu and Tao Zhe and Dongjie Wang and Zijun Yao and
               Kunpeng Liu and Yanjie Fu and Huan Liu and Jian Pei},
  title     = {{AgentOS}: From Application Silos to a Natural Language-Driven
               Data Ecosystem},
  journal   = {arXiv preprint arXiv:2603.08938},
  year      = {2026}
}

@inproceedings{liu2024repobench,
  author    = {Tianyang Liu and Canwen Xu and Julian J. McAuley},
  title     = {{RepoBench}: Benchmarking Repository-Level Code Auto-Completion
               Systems},
  booktitle = {The Twelfth International Conference on Learning Representations
               (ICLR 2024)},
  year      = {2024}
}

@inproceedings{liu2024cachegen,
  author    = {Yuhan Liu and Hanchen Li and Yihua Cheng and Siddhant Ray and
               Yuyang Huang and Qizheng Zhang and Kuntai Du and Jiayi Yao and
               Shan Lu and Ganesh Ananthanarayanan and Michael Maire and
               Henry Hoffmann and Ari Holtzman and Junchen Jiang},
  title     = {{CacheGen}: {KV} Cache Compression and Streaming for Fast Large
               Language Model Serving},
  booktitle = {Proceedings of the ACM SIGCOMM 2024 Conference
               (SIGCOMM 2024)},
  pages     = {38--56},
  year      = {2024}
}

@article{liu2025lmcache,
  author    = {Yuhan Liu and Jiayi Yao and Yihua Cheng and Yuwei An and
               Xiaokun Chen and Shaoting Feng and Yuyang Huang and Samuel Shen and
               Rui Zhang and Kuntai Du and Junchen Jiang},
  title     = {{LMCache}: An Efficient {KV} Cache Layer for Enterprise-Scale
               {LLM} Inference},
  journal   = {arXiv preprint arXiv:2510.09665},
  year      = {2025}
}

@article{autellix,
  author    = {Michael Luo and Xiaoxiang Shi and Colin Cai and Tianjun Zhang and
               Justin Wong and Yichuan Wang and Chi Wang and Yanping Huang and
               Zhifeng Chen and Joseph E. Gonzalez and Ion Stoica},
  title     = {{Autellix}: An Efficient Serving Engine for {LLM} Agents as
               General Programs},
  journal   = {arXiv preprint arXiv:2502.13965},
  year      = {2025}
}

@inproceedings{aios,
  author    = {Kai Mei and Xi Zhu and Wujiang Xu and Mingyu Jin and Wenyue Hua and
               Zelong Li and Shuyuan Xu and Ruosong Ye and Yingqiang Ge and
               Yongfeng Zhang},
  title     = {{AIOS}: {LLM} Agent Operating System},
  booktitle = {Second Conference on Language Modeling (COLM 2025)},
  year      = {2025}
}

@inproceedings{merrill2026terminalbench,
  author    = {Mike A. Merrill and Alexander Glenn Shaw and Nicholas Carlini and
               Boxuan Li and Harsh Raj and Ivan Bercovich and Lin Shi and
               Jeong Yeon Shin and Thomas Walshe and E. Kelly Buchanan and
               Junhong Shen and Guanghao Ye and Haowei Lin and Jason Poulos and
               Maoyu Wang and Jenia Jitsev and Marianna Nezhurina and Di Lu and
               Orfeas Menis Mastromichalakis and Zhiwei Xu and Zizhao Chen and
               Yue Liu and Robert Zhang and Leon Liangyu Chen and Anurag Kashyap and
               Jan-Lucas Uslu and Jeffrey Li and Jianbo Wu and Minghao Yan and
               Song Bian and Vedang Sharma and Ke Sun and Steven Dillmann and
               Akshay Anand and Andrew Lanpouthakoun and Bardia Koopah and
               Changran Hu and Etash Kumar Guha and Gabriel H. Dreiman and
               Jiacheng Zhu and Karl Krauth and Li Zhong and Niklas Muennighoff and
               Robert Kwesi Amanfu and Shangyin Tan and Shreyas Pimpalgaonkar and
               Tushar Aggarwal and Xiangning Lin and Xin Lan and Xuandong Zhao and
               Yiqing Liang and Yuanli Wang and Zilong Wang and Changzhi Zhou and
               David Heineman and Hange Liu and Harsh Trivedi and John Yang and
               Junhong Lin and Manish Shetty and Michael Yang and Nabil Omi and
               Negin Raoof and Shanda Li and Terry Yue Zhuo and Wuwei Lin and
               Yiwei Dai and Yuxin Wang and Wenhao Chai and Shang Zhou and
               Dariush Wahdany and Ziyu She and Jiaming Hu and Zhikang Dong and
               Yuxuan Zhu and Sasha Cui and Ahson Saiyed and
               Arinbj{\"o}rn Kolbeinsson and Christopher Michael Rytting and
               Ryan Marten and Yixin Wang and Alex Dimakis and Andy Konwinski and
               Ludwig Schmidt},
  title     = {{Terminal-Bench}: Benchmarking Agents on Hard, Realistic Tasks
               in Command Line Interfaces},
  booktitle = {The Fourteenth International Conference on Learning Representations
               (ICLR 2026)},
  year      = {2026}
}

@inproceedings{miao2024spotserve,
  author    = {Xupeng Miao and Chunan Shi and Jiangfei Duan and Xiaoli Xi and
               Dahua Lin and Bin Cui and Zhihao Jia},
  title     = {{SpotServe}: Serving Generative Large Language Models on
               Preemptible Instances},
  booktitle = {Proceedings of the 29th ACM International Conference on
               Architectural Support for Programming Languages and Operating
               Systems (ASPLOS 2024), Volume 2},
  pages     = {1112--1127},
  year      = {2024}
}

@article{ni2025gittaskbench,
  author    = {Ziyi Ni and Huacan Wang and Shuo Zhang and Shuo Lu and
               Ziyang He and Wang You and Zhenheng Tang and Sen Hu and Bo Li and
               Chen Hu and Binxing Jiao and Daxin Jiang and Yuntao Du and Pin Lyu},
  title     = {{GitTaskBench}: A Benchmark for Code Agents Solving Real-World
               Tasks Through Code Repository Leveraging},
  journal   = {Proceedings of the AAAI Conference on Artificial Intelligence
               (AAAI 2026)},
  volume    = {40},
  number    = {38},
  pages     = {32564--32572},
  year      = {2026}
}

@misc{openaiDeepResearch2025,
  author    = {{OpenAI}},
  title     = {Introducing {Deep Research}},
  year      = {2025},
  url       = {https://openai.com/index/introducing-deep-research/}
}

@misc{openai_gpt_oss_2025,
  author    = {{OpenAI}},
  title     = {{gpt-oss-120b} \& {gpt-oss-20b} Model Card},
  year      = {2025},
  url       = {https://openai.com/index/gpt-oss-model-card/}
}

@misc{openaiCodexPrompting2026,
  author    = {{OpenAI}},
  title     = {Introducing Codex},
  year      = {2025},
  url       = {https://openai.com/index/introducing-codex/}
}

@article{memgpt,
  author    = {Charles Packer and Sarah Wooders and Kevin Lin and Vivian Fang and
               Shishir G. Patil and Ion Stoica and Joseph E. Gonzalez},
  title     = {{MemGPT}: Towards {LLMs} as Operating Systems},
  journal   = {arXiv preprint arXiv:2310.08560},
  year      = {2023}
}

@inproceedings{alto,
  author    = {Keshav Santhanam and Deepti Raghavan and Muhammad Shahir Rahman and
               Thejas Venkatesh and Neha Kunjal and Pratiksha Thaker and
               Philip Levis and Matei Zaharia},
  title     = {{ALTO}: An Efficient Network Orchestrator for Compound {AI} Systems},
  booktitle = {Proceedings of the 4th Workshop on Machine Learning and Systems
               (EuroMLSys 2024)},
  pages     = {117--125},
  year      = {2024}
}

@inproceedings{song2024powerinfer,
  author    = {Yixin Song and Zeyu Mi and Haotong Xie and Haibo Chen},
  title     = {{PowerInfer}: Fast Large Language Model Serving with a
               Consumer-Grade {GPU}},
  booktitle = {Proceedings of the ACM SIGOPS 30th Symposium on Operating
               Systems Principles (SOSP 2024)},
  pages     = {590--606},
  year      = {2024}
}

@inproceedings{llumnix,
  author    = {Biao Sun and Ziming Huang and Hanyu Zhao and Wencong Xiao and
               Xinyi Zhang and Yong Li and Wei Lin},
  title     = {Llumnix: Dynamic Scheduling for Large Language Model Serving},
  booktitle = {18th USENIX Symposium on Operating Systems Design and
               Implementation (OSDI 2024)},
  pages     = {173--191},
  year      = {2024}
}

@inproceedings{ayo,
  author    = {Xin Tan and Yimin Jiang and Yitao Yang and Hong Xu},
  title     = {Towards End-to-End Optimization of {LLM}-based Applications
               with {Ayo}},
  booktitle = {Proceedings of the 30th ACM International Conference on
               Architectural Support for Programming Languages and Operating
               Systems (ASPLOS 2025), Volume 2},
  pages     = {1302--1316},
  year      = {2025}
}

@inproceedings{wang2025openhands,
  author    = {Xingyao Wang and Boxuan Li and Yufan Song and Frank F. Xu and
               Xiangru Tang and Mingchen Zhuge and Jiayi Pan and Yueqi Song and
               Bowen Li and Jaskirat Singh and Hoang H. Tran and Fuqiang Li and
               Ren Ma and Mingzhang Zheng and Bill Qian and Yanjun Shao and
               Niklas Muennighoff and Yizhe Zhang and Binyuan Hui and Junyang Lin and
               Robert Brennan and Hao Peng and Heng Ji and Graham Neubig},
  title     = {{OpenHands}: An Open Platform for {AI} Software Developers as
               Generalist Agents},
  booktitle = {The Thirteenth International Conference on Learning Representations
               (ICLR 2025)},
  year      = {2025}
}

@inproceedings{wu2024autogen,
  author    = {Qingyun Wu and Gagan Bansal and Jieyu Zhang and Yiran Wu and
               Beibin Li and Erkang Zhu and Li Jiang and Xiaoyun Zhang and
               Shaokun Zhang and Jiale Liu and Ahmed Hassan Awadallah and
               Ryen W. White and Doug Burger and Chi Wang},
  title     = {{AutoGen}: Enabling Next-Gen {LLM} Applications via Multi-Agent
               Conversations},
  booktitle = {First Conference on Language Modeling (COLM 2024)},
  year      = {2024}
}

@inproceedings{xie2024osworld,
  author    = {Tianbao Xie and Danyang Zhang and Jixuan Chen and Xiaochuan Li and
               Siheng Zhao and Ruisheng Cao and Toh Jing Hua and Zhoujun Cheng and
               Dongchan Shin and Fangyu Lei and Yitao Liu and Yiheng Xu and
               Shuyan Zhou and Silvio Savarese and Caiming Xiong and
               Victor Zhong and Tao Yu},
  title     = {{OSWorld}: Benchmarking Multimodal Agents for Open-Ended Tasks in
               Real Computer Environments},
  booktitle = {Advances in Neural Information Processing Systems
               (NeurIPS 2024)},
  volume    = {37},
  pages     = {52040--52094},
  year      = {2024}
}

@article{qwen3,
  author    = {An Yang and Anfeng Li and Baosong Yang and Beichen Zhang and
               Binyuan Hui and Bo Zheng and Bowen Yu and Chang Gao and
               Chengen Huang and Chenxu Lv and Chujie Zheng and Dayiheng Liu and
               Fan Zhou and Fei Huang and Feng Hu and Hao Ge and Haoran Wei and
               Huan Lin and Jialong Tang and Jian Yang and Jianhong Tu and
               Jianwei Zhang and Jianxin Yang and Jiaxi Yang and Jing Zhou and
               Jingren Zhou and Junyang Lin and Kai Dang and Keqin Bao and
               Kexin Yang and Le Yu and Lianghao Deng and Mei Li and Mingfeng Xue and
               Mingze Li and Pei Zhang and Peng Wang and Qin Zhu and Rui Men and
               Ruize Gao and Shixuan Liu and Shuang Luo and Tianhao Li and
               Tianyi Tang and Wenbiao Yin and Xingzhang Ren and Xinyu Wang and
               Xinyu Zhang and Xuancheng Ren and Yang Fan and Yang Su and
               Yichang Zhang and Yinger Zhang and Yu Wan and Yuqiong Liu and
               Zekun Wang and Zeyu Cui and Zhenru Zhang and Zhipeng Zhou and
               Zihan Qiu},
  title     = {{Qwen3} Technical Report},
  journal   = {arXiv preprint arXiv:2505.09388},
  year      = {2025}
}

@inproceedings{yang2024sweagent,
  author    = {John Yang and Carlos E. Jimenez and Alexander Wettig and
               Kilian Lieret and Shunyu Yao and Karthik Narasimhan and
               Ofir Press},
  title     = {{SWE-agent}: Agent-Computer Interfaces Enable Automated
               Software Engineering},
  booktitle = {Advances in Neural Information Processing Systems
               (NeurIPS 2024)},
  volume    = {37},
  pages     = {50528--50652},
  year      = {2024}
}

@inproceedings{yao2024tot,
  author    = {Shunyu Yao and Dian Yu and Jeffrey Zhao and Izhak Shafran and
               Thomas L. Griffiths and Yuan Cao and Karthik Narasimhan},
  title     = {Tree of Thoughts: Deliberate Problem Solving with Large
               Language Models},
  booktitle = {Advances in Neural Information Processing Systems
               (NeurIPS 2023)},
  volume    = {36},
  pages     = {11809--11822},
  year      = {2023}
}

@inproceedings{yu2022orca,
  author    = {Gyeong-In Yu and Joo Seong Jeong and Geon-Woo Kim and
               Soojeong Kim and Byung-Gon Chun},
  title     = {Orca: A Distributed Serving System for {Transformer-Based}
               Generative Models},
  booktitle = {16th USENIX Symposium on Operating Systems Design and
               Implementation (OSDI 2022)},
  pages     = {521--538},
  year      = {2022}
}

@inproceedings{zhang2024inftybench,
  author    = {Xinrong Zhang and Yingfa Chen and Shengding Hu and Zihang Xu and
               Junhao Chen and Moo Hao and Xu Han and Zhen Thai and Shuo Wang and
               Zhiyuan Liu and Maosong Sun},
  title     = {$\infty${Bench}: Extending Long Context Evaluation Beyond
               100{K} Tokens},
  booktitle = {Proceedings of the 62nd Annual Meeting of the Association for
               Computational Linguistics (ACL 2024), Volume 1: Long Papers},
  pages     = {15262--15277},
  year      = {2024}
}

@inproceedings{zheng2024sglang,
  author    = {Lianmin Zheng and Liangsheng Yin and Zhiqiang Xie and Chuyue Sun and
               Jeff Huang and Cody Hao Yu and Shiyi Cao and Christos Kozyrakis and
               Ion Stoica and Joseph E. Gonzalez and Clark Barrett and Ying Sheng},
  title     = {{SGLang}: Efficient Execution of Structured Language Model Programs},
  booktitle = {Advances in Neural Information Processing Systems
               (NeurIPS 2024)},
  volume    = {37},
  pages     = {62557--62583},
  year      = {2024}
}

@inproceedings{zhang2026jitserve,
  author    = {Wei Zhang and Zhiyu Wu and Yi Mu and Rui Ning and Banruo Liu and
               Nikhil Sarda and Myungjin Lee and Fan Lai},
  title     = {{JITServe}: {SLO-aware} {LLM} Serving with Imprecise Request
               Information},
  booktitle = {23rd USENIX Symposium on Networked Systems Design and
               Implementation (NSDI 2026)},
  pages     = {825--848},
  year      = {2026}
}

@misc{opencode2025,
  author    = {{Anomaly}},
  title     = {{OpenCode}: The Open Source Coding Agent},
  year      = {2025},
  url       = {https://github.com/anomalyco/opencode}
}

@article{bian2025tokencake,
  author    = {Zhuohang Bian and Feiyang Wu and Zhuoran Li and Teng Ma and
               Youwei Zhuo},
  title     = {{TokenCake}: A {KV}-Cache-Centric Serving Framework for
               {LLM}-Based Multi-Agent Applications},
  journal   = {arXiv preprint arXiv:2510.18586},
  year      = {2025}
}
}

\end{document}